\documentclass[pss]{wiley2sp} 
\usepackage{braket}						   
\usepackage{blindtext}
\usepackage{bm}
\usepackage{amsmath,amssymb,amsfonts,latexsym,times}
\usepackage[english]{babel}
\usepackage{graphicx}
\usepackage{xcolor}
\usepackage{color}
\usepackage[pdftex]{hyperref}
\usepackage{enumitem}
\usepackage{multicol}

\begin{document}

\title{A scalable numerical approach to the solution of the Dyson equation for the non-equilibrium single-particle Green's function}

\author{%
  W. Talarico\textsuperscript{\Ast,\textsf{\bfseries 1}},
  S. Maniscalco\textsuperscript{\textsf{\bfseries 1,2}},
  N. Lo Gullo\textsuperscript{\textsf{\bfseries 1}}}

\mail{e-mail
  \textsf{nawata@utu.fi}}

\institute{%
  \textsuperscript{1}\,QTF Centre of Excellence, Turku Centre for Quantum Physics, Department of Physics and 
Astronomy, University of Turku, 20014 Turku, Finland\\
  \textsuperscript{2}\, QTF Centre of Excellence, Department of Applied Physics, Aalto University, FI-00076 Aalto, Finland}


\abstract{\bf%
  In this work we present a numerical method to solve the set of Dyson-like equations arising the context of non-equilibrium Green's functions.
  The technique is based on the self-consistent solution of the Dyson equations for the interacting single-particle Green's function once a choice for the self-energy, functional of the single-particle Green's function itself, is done. We briefly review the theory of the non-equilibrium Green's functions in order to highlight 
  the main point useful in discussing the proposed technique. We also discuss the relation between our approach and the textbook approach to solve 
  the Kadanoff-Baym equations.
  We then present and discuss the numerical implementation which is based on the distribution of the elements of the Green's function and self-energies on
  a grid of processes. We discuss how the structure of the considered self-energy approximations influences the distribution of the matrices in order to minimize
  the communication time among processes and which should be considered in the case of other approximations.
  We give an example of the application of our technique to the case of quenches in ultracold gases, also discussing to the convergence features of our 
  scheme.
  }

\maketitle   

\selectlanguage{english}

\section{Introduction}
In recent years, there has been a growing interest in investigating the dynamics of 
many-body quantum systems triggered by technological developments in different fields which allowed 
the realization of setups such as the ultracold-gases, or the pump and probe techniques applied to solid state systems 
where dynamical properties can be resolved and controlled to a very good extent.
One representative example is that of the equilibration of a closed system 
following a sudden and short external perturbation; the theoretical efforts in this context
has led to the discovery of the so called many-body localization (MBL) 	\cite{greiner,arijeet,bardarson,serbyn}, a phenomenon which is still 
highly debated in the field of condensed matter physics.
It is also worthwhile to mention the discovery of out-of-equilibrium phases of matter induced in
many-body system by external perturbation such as the light-induced superconductivity \cite{fausti}
and/or gap engineering in semiconductors \cite{kandelaki}.

The intense experimental research activity has contributed to an equally intense search for theoretical approaches 
able to capture such features. The available theoretical tools are mostly still in their infancy or lacking altogether.
Understanding those phenomena unavoidably requires a framework which can take into account
effects of correlations in a many-body quantum system with possibly time-dependent (TD) external perturbations.
The starting point is the TD Schreodinger equation (SE) whose numerical solution
is computationally challenging already in a very small reduced subspace of many-electrons states.
Therefore several time-propagation algorithms with different degree of approximation 
and range of applicability have been developed.
Some of these techniques include: the time-dependent density matrix renormalization group (TD-DMRG) \cite{tddmrg}, 
time-dependent numerical renormalization group (TD-NRG) \cite{nrg1,nrg2}, 
functional renormalization group (FRG) \cite{frg} \cite{frg1} \cite{frg2}, 
diagrammatic many-body methods \cite{diagram} \cite{diagram1}, 
and Quantum Monte Carlo (QMC) techniques \cite{qmc} \cite{qmc1}.
Lately, strongly correlated systems have also been studied with density functional theory (DFT) 
both in its static [\cite{dft}-\cite{dft4}] and time-dependent version [\cite{tddft}-\cite{tddft5}].
Even though each of these techniques has given good results in its own range of applicability
in describing the equilibration and steady-state regime, 
each of these techniques are limited in different degree of approximation. 
As an example, although TD-DFT is an exact reformulation of the TD-SE, 
in practice all simulations lack of dynamical exchange-correlation effects 
which often play a major role in, e.g., charge migration processes.\\   

A very powerful approach, which is the subject of our work, 
is that of the Non-Equilibrium Green's Functions (NEGFs)
which has been developed at the same time as their well know equilibrium counterpart.
One of its most important features is that it allows to retain 
the formal structure of the many-body perturbation theory (MBPT) developed for the equilibrium case,
therefore allowing to extend all known perturbation schemes to the non-equilibrium scenario.
A drawback of this approach is that the system of equations needed to be solved
in order to have access to information on physical quantities, namely the Kadanoff-Baym equations (KBEs), 
is hopelessly complicated for an analytic approach and it is very demanding for any numerical approach,
especially for computers available at the time when this technique was developed.
The reason lies on the two-real-time structure (plus possibly one extra imaginary time axis if considering an initially
interacting state) of the single-particle Green's function.
This resulted in the nearly total abandonment of the application and development
of NEGFs approach up until very recently.
In more recent years, with the advent of supercomputers, 
the interest in the NEGFs has witnessed a renewed attention
with the development of numerical techniques for the solution of the KBEs~\cite{stan,bonitz}
which, although still computationally very demanding, has found several applications in the study of
out-of-equilibrium interacting many-body systems~\cite{stef van,balzer}.
In order to lower the computational load and therefore to investigate longer time dynamics, 
a simplified version of the KBEs is often employed. The essence of this approximation, introduced in Ref.~\cite{lipavsky},
is to compute two-times functions in a simplified way by means of the so-called Generalized-Kadanoff-Baym-Ansatz (GKBA),
which in practice amounts to calculate the spectrum of the system (the Retarded/Advanced Green's functions) in a way which does not require
the two-times propagation, typically at the Hartree-Fock level. 
The result of applying this procedure to the original KBEs is to return an integro-differential equation for the
single-particle density matrix which can be solved using standard numerical techniques for such class of equations.

This approach is particularly powerful as it is completely general and can be applied to different quantum systems,
both closed and open as already shown in the original paper. 
On the other hand, the application of such approach to different cases has shown that 
it suffers from two major drawbacks: 1) the single particle spectrum is
often calculated only at the Hartree-Fock regime; 2) the initial state of the system has to be chosen uncorrelated (non-interacting).
A solution to this last problem has been recently proposed exploiting an analogy between the equilibrium situation
and the non-equilibrium one by D. Karlsson and collaborators in Ref.\cite{Karlsson2018}.
On the other hand, the validity of such approximation to different systems is still under scrutiny together with
new methods to improve it.
For this reason, it is often necessary to compare its results with the solution of the full KBEs in order to 
understand the range of validity of the GKBA-master equation and whether its limitations come from
an incorrect evaluation of the spectral function (thus from how correlations in the system are accounted for) 
or from neglecting the initially correlated state of the system which in the KBEs can be easily included.

In our work we propose a new approach for the solution of the KBEs starting from a slightly different,
yet completely equivalent, point of view.
The problem of finding the evolution of the single particle Green's function for an interacting many-body system
can be formulated either in the framework of a system of coupled equations of motion
for the n-particle Green’s functions (the Martin-Swinger hierarchy), 
or as a series expansion of the evolution operator giving different contributions
which are then resummed to obtain a Dyson equation.
Actually in this latter case the one ends up with the solution of a set of five integral equations: the Hedin equations~\cite{stef van}.
In this paper we propose a numerical approach to the solution of the Dyson equation for the single particle
Green's functions for an interacting system possibly coupled to external reservoirs.
Our approach is totally equivalent to the solution of the KBE and allows for the 
inclusion of dynamical correlations in a self-consistent manner; 
at the same time it shares with TD-DFT the scaling of the computational cost with the system size. 

In what follows we will focus on the case of fermionic particles, 
but of course the extension of the formalism to bosonic particles is straightforward.
One example of the application of our method to the bosonic case can is discussed in Ref.~\cite{LoGullo2016}
where a quench in one- and two-dimensional Bose-Hubbard model has been studied.
Other types of systems can be also covered by the formalism presented here as for instance 
the case of phonons or of any other type of quasi-particle. 
The key difference between these cases and the approach presented hereafter 
will mostly be in the definition of the single particle Green's functions and, 
in turn, of the self-energies which embody the main features of interactions.

\section{General framework}
In this and the following section we review for completeness the key ingredients 
of the non-equilibrium Green's function theory and of the main approximations used to treat interactions
in weakly interacting fermions and the coupling of such a system to external reservoirs to describe transport.

\subsection{Closed interacting system}

We consider a many-body system whose particles interact via a two-body potential.
Our aim is to describe the time evolution of the system when the latter is brought 
out-of-equilibrium by a time-dependent external potential or it is subject to a change in some of the system's parameters (quantum quench).
This external field could be a bias voltage in a quantum transport case or the change in the trapping potential in the case of ultracold gases; 
in this latter case, a sudden change of particle-particle interactions can also be realized.
In second quantization the time-dependent Hamiltonian 
of a system of interacting identical particles reads

\begin{align}
&\hat{H}_C(t) = \int d \bold{x} \hat{\psi}^\dagger (\bold{x}) h(\bold{x},t) \hat{\psi} ( \bold{x}) \nonumber \\
&+\frac{1}{2}\int \int d \bold{x} d \bold{x'} \hat{\psi}^\dagger (\bold{x}) \hat{\psi}^\dagger (\bold{x'}) v(\bold{r}, \bold{r'},t) \hat{\psi} (\bold{x}) \hat{\psi}, (\bold{x'})
\end{align}
where the field operator $\hat{\psi}$ ( $\hat{\psi}^\dagger$) 
with argument $\bold{x}= (\bold{r},\sigma) $ annihilates (creates) a fermion 
in position $\bold{r}$ with spin $\sigma$. 
The two-body interaction are given by the function  $v(\bold{r}, \bold{r'})$ 
whose strength typically depends on the distance $|\bold{r}-\bold{r'}|$, 
like in the Coulomb repulsion. We use the subscript $C$ in the Hamiltonian 
to distinguish it in the context of a more general framework where the system is coupled to external reservoirs
and it therefore becomes the "central" region of a larger system which includes the reservoirs themselves (see next subsection).
The one-body part of the Hamiltonian is

\begin{equation}
h(\bold{x},t) = -\frac{1}{2} \nabla^2 + w(\bold{x},t) -\mu
\end{equation}
where $w(\bold{x},t)$ is a time-dependent external field and $\mu$ 
the chemical potential of the initial equilibrium system. \\ 
If we restrict ourselves to a suitable basis representation, 
like the spin-orbital basis $ \varphi_{i \sigma}(\bold{x})=\varphi_i(\bold{r}) \delta_{\sigma \sigma'}$, 
then we can define creation and annihilation operators $\hat{d}_{i \sigma}^\dagger,\hat{d}_{i \sigma}$ 
for the state $i \sigma$ as a linear combination of field operators 
at different position-spin coordinates 

\begin{align}
\label{ddag}
&\hat{d}_{i \sigma}^\dagger \equiv \int d \bold{x} \;\varphi_{i \sigma} (\bold{x}) \hat{\psi}^\dagger (\bold{x}) \\ 
\label{d}
&\hat{d}_{i \sigma} \equiv \int d \bold{x} \;\varphi_{i \sigma}^* (\bold{x}) \hat{\psi} (\bold{x}). 
\end{align} 

where the integral is $\int d \bold{x} = \sum_{\sigma} \int d \bold{r} $.
The operators  $\hat{d}_{i \sigma}^\dagger $ and $\hat{d}_{i \sigma}$ 
inherit the anti-commutation rules from the field operators $\hat{\psi}^\dagger $ and $\hat{\psi} $.
Eq. \eqref{ddag}, \eqref{d}, together with their anti-commutation relations, 
leads to the following representation for the Hamiltonian 

\begin{align}
\label{intsys}
\hat{H}_C (t) &= \hat{H}_0 (t) + \hat{V} (t)\\
\hat{H}_0 (t) &= \sum_{ij, \sigma} h_{ij} (t) \hat{d}_{i \sigma}^\dagger \hat{d}_{j \sigma} \\
\hat{V} (t)  &= \frac{1}{2} \sum_{ijkl} \sum_{\sigma \sigma'} v_{ijkl}(t) \hat{d}_{i \sigma}^\dagger \hat{d}_{j \sigma'}^\dagger \hat{d}_{k \sigma'} \hat{d}_{l \sigma}  
\end{align} 
with the single-particle Hamiltonian matrix given by 

\begin{equation}
h_{ij}(t)=  \int d \bold{r} \;\varphi_{i}^* (\bold{r}) h(\bold{r},t) \varphi_{j} (\bold{r})
\end{equation} 
and the so called Coulomb integral

\begin{equation}
\label{coulomb}
 v_{ijkl}(t) = \int d \bold{r} d \bold{r'}\; \varphi_{i}^* (\bold{r}) \varphi_{j}^* (\bold{r'}) v(\bold{r}, \bold{r'},t) \varphi_{k} (\bold{r'}) \varphi_{l} (\bold{r}).
\end{equation}
The above expressions can be generalized in case of the presence of a magnetic field and/or spin
orbit coupling.

\subsection{Coupling to external reservoirs}

To study quantum transport through a correlated system  
we connect a central interacting open quantum system described by Eq.~\ref{intsys} 
to non-interacting electronic reservoirs.
The model Hamiltonian for such a scenario becomes then:
 
\begin{equation}
\label{totalH}
\hat{H} (t) =  \hat{H}_C (t) + \hat{H}_{leads} (t) + \hat{H}_T 
\end{equation}

where $\hat{H}_C (t)$ is the Hamiltonian of the central region Eq. \eqref{intsys}, 
the infinite leads are described by the terms $\hat{H}_{leads} (t)= \sum_\alpha \hat{H}_{\alpha} (t)$ 
whose second quantized form reads 

\begin{equation}
\hat{H}_{leads} (t)= \sum_{\alpha} \sum_{ij,\sigma} [t_{ij}^\alpha + V_\alpha(t)\delta_{ij}] \hat{c}_{ \alpha i \sigma}^\dagger \hat{c}_{\alpha i \sigma}
\end{equation}

here the operators $ \hat{c}^\dagger $, $\hat{c}$ are the creation and annihilation operators 
for the reservoir $\alpha$, $t_{ij}^\alpha $ is 
the nearest neighbor Hamiltonian and $V_\alpha (t)$ denotes the time-dependent bias for each lead.
The last term $\hat{H}_T $ describes the contact of the correlated system to the electrodes, 
i.e., the hybridization of the central region with the levels of the leads:

\begin{align}
\label{cplngH}
\hat{H}_T & = \sum_{\alpha} \sum_{ij,\sigma}  [T_{i,j \alpha} \hat{d}_{i \sigma}^\dagger \hat{c}_{\alpha j \sigma} + T_{\alpha j ,i}^* \hat{c}_{\alpha j \sigma}^\dagger \hat{d}_{i \sigma} ]\\
 &=\sum_{\alpha} \hat{H}_T^\alpha (t)\nonumber,
\end{align}
where $ T_{\alpha i,j}$ are the matrix elements of the coupling between the leads $\alpha$
and the system. \\

\subsection{Single particle Green's functions}

\begin{figure}[t]
\begin{center}
\includegraphics[width=.3\textwidth]{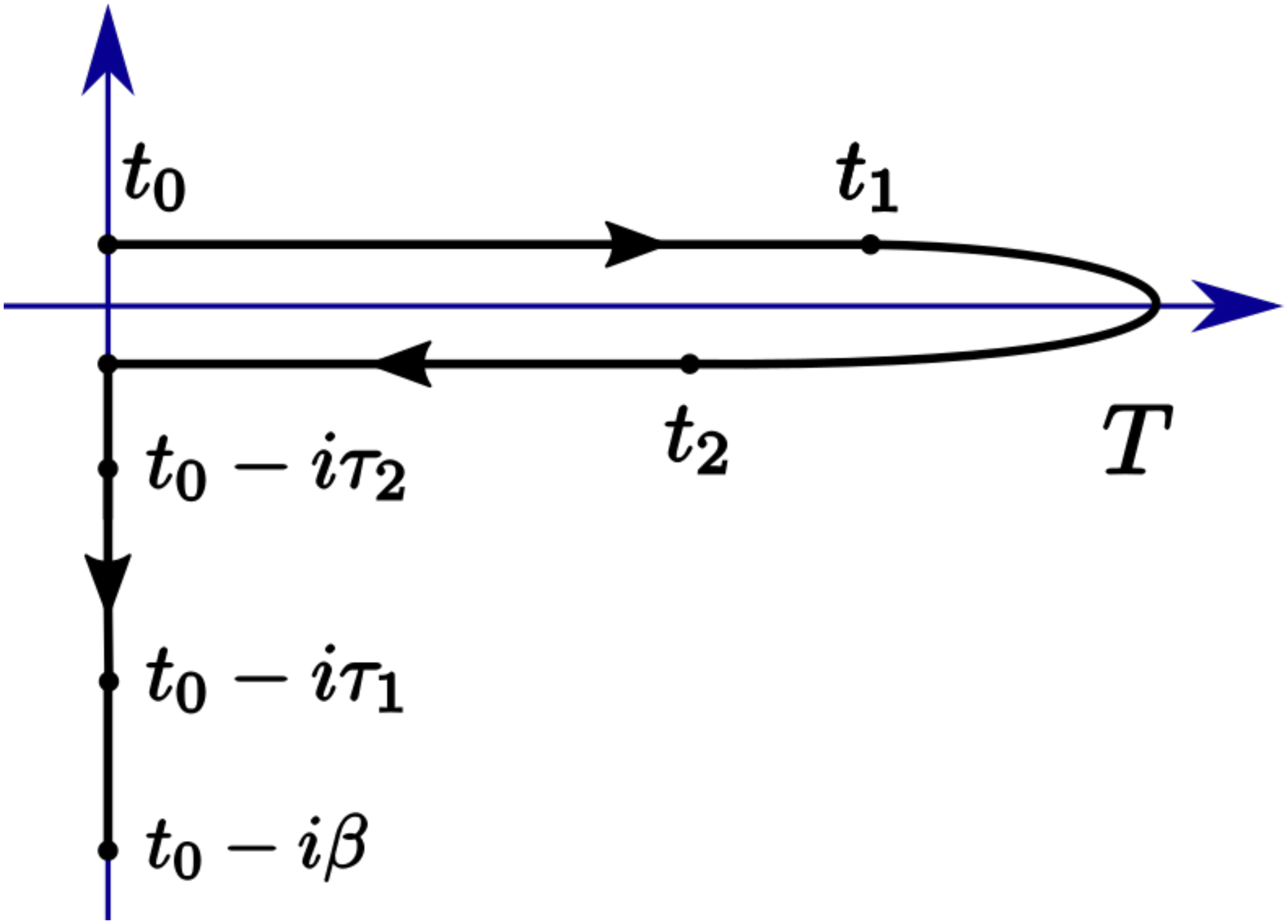}
\caption{The Keldysh-Schwinger contour $\gamma$. The arrows in the contour indicates the 
ordering of the arguments along the contour defining the relative order of the complex times, 
namely $t_0>_{\gamma}t_1>_{\gamma}t_2>_{\gamma}t_0-i \tau_2>_{\gamma}t_0-i\beta>_{\gamma}$.}
\label{figkeldysh}
\end{center}
\end{figure}

The starting point in the theory of the non-equilibrium Green's function is the definition
of the single-particle Green's function as 
the expectation value of the contour-ordered product 
of the creation and annihilation operators 
\begin{equation}
\label{green}
G(1;1') = -i \left\langle \mathcal{T}_\gamma \left[ \hat{\psi}_H(1) \hat{\psi}_H^\dagger (1') \right] \right\rangle _0
\end{equation}
where the subscript $H$ denotes the Heisenberg picture 
and the indexes $1={\bold{x}_{1},z_1}$ and $1'={\bold{x}_{1}',z_{1}'}$ are collective indexes for position, spin and complex-time. 
Furthermore, $z$ and $z'$ are a contour time variables and $\mathcal{T}_\gamma$ 
orders the operators along the Keldysh contour $\gamma$ by arranging 
the operators with later contour times to left Fig. \ref{figkeldysh}. 
Similarly, we denote by $h(1)$ the matrix elements of the first quantized Hamiltonian $\hat{h}$
in the position, spin and complex-time indexes.
Here the symbol $\left\langle \dots \right\rangle _0$ denotes the average 
over the initial many-body thermal state. 
By applying $\left[ i \partial_{z_1} - h(z) \right] $ and using the Heisenberg equations of motion for the 
operators $\hat{\psi}$ and $\hat{\psi}^\dagger$ under the evolution given by $\hat{H}_C(t)$,
one obtains the first equation of the Martin-Schwinger hierarchy (MSH):
\begin{align}
\label{oneMS}
 & \left[ i \partial_{z_1} - h(1) \right]  G(1 ; 1') = \delta(1, 1')  \\
 &- \int d \bar{1}  v(1,\bar{1})  G_2 (1,\bar{1}; 1',\bar{1}^+) \nonumber
\end{align}
where the two-particle Green's function $G_2 (1,2; 1',2')=(-i)^2 \left\langle \mathcal{T}_\gamma \left[ \hat{\psi}_H (1) \hat{\psi}_H (2) \hat{\psi}_H^\dagger (2')\hat{\psi}_H^\dagger (1') \right] \right\rangle _0$ has been introduced.
Here the integral $\int d \bar{1} = \int d \bar{\bold{x}}_1 \int_{\gamma} d \bar{z}_1$ and $1^+= \bold{x}_1, z_1 +\delta$ denoting a time infinitesimally larger than $z_1$ on the Keldysh contour $\gamma$.
A similar equation is obtained by acting to the left with the operator $\left[ -i \partial_{z_1'} - h(1') \right] $.
In order to truncate the MSH, a self-energy $\Sigma$ is introduced in such a way that:
\begin{align}
\label{SEMS}
 & \int d \bar{1}\Sigma(1;\bar{1})  G(\bar{1} ; 1') = \\
 &- \int d \bar{1}  v(1,\bar{1})  G_2 (1,\bar{1}; 1',\bar{1}^+) \nonumber
\end{align}
The physical meaning of the self-energy $\Sigma$ is to introduce an effective function
which accounts for the two particles scattering which are encoded into the two-particle Green's function $G_2$.
Thanks to the definition of $\Sigma$ with Eq.~\ref{SEMS} one obtains the following equation
on the complex contour for the single particle Green's functions:
\begin{align}
\label{KBeq}
& \left[ i \partial_{z_1} - h(1) \right]  G(1;1') = \delta(1,1')+ \int d \bar{1} \Sigma(1, \bar{1})  G (\bar{1};1') \nonumber \\
\end{align}
which has to be solved with the Kubo-Martin-Schwinger (KMS) boundary conditions $G(t_0, z_1') = - G(t_0 - i \beta , z_1')$, 
following directly from \eqref{green} and the cyclic property of the trace.    
By means of the Langreth rules, it is possible to project these equations of motion 
for the single-particle Green's function onto the real and imaginary axis.
The resulting set of equations for those components are called the Kadanoff-Baym equations and represent, 
together with the initial conditions, the usual way to completely determine
the single-particle Green's functions once a choice for the self-energy is made.
The numerical implementation of the solution of such equations requires a two-times propagation
scheme~\cite{stef van,stan,bonitz,balzer}.

An alternative approach to find the interacting Green's function in Eq.~\ref{green}
is to go to the interaction picture and then expand the evolution operator containing the interaction
term $\hat{V}_I(t)$, where the subscript $I$ stands for "interaction picture" and then use Wick's theorem to 
write down a series expansion of the interacting single-particle Green's function in terms of the non-interacting one 
$g_0 (1;1')$ which satify the following equations:

\begin{align}
\label{nnintgreen}
&\left[ i \partial_{z_1} - h(1) \right] g_0(1; 1') = \delta(1, 1')\\
&g_0(1; 1')[ -i\overleftarrow{\partial}_{z_1'} - h(1') ]= \delta(1, 1').
\end{align}

By collecting the wanted terms of this series expansion into a self-energy term,
the Dyson equation is obtained:
\begin{align}
\label{dysongreen}
 G(1;1') &= g_0 (1;1') + \int d \bar{1} d \bar{2} g_0 (1;\bar{1}) \Sigma(\bar{1}, \bar{2}) G(\bar{2}, 1')
\end{align}
The Dyson equation is the formal solution 
of the Martin-Schwinger hierarchy for the one-particle Green's function
as can be easily seen by applying $\left[ i  \delta(1,1') \partial_{z_1} - h(1',1) \right]$ and integrating over $1$ 
to Eq.\eqref{dysongreen} to obtain Eq.~\eqref{KBeq}.

Therefore solving both approaches are fully equivalent
in describing the time propagation of the full non-equilibrium Green's function.
In the following we derive a procedure for the self-consistent solution of the Dyson equation.\\

Before proceeding, we want to recall the main definitions
for function with time indexes on the real and imaginary branches of the contour.
Any two-time function defined on the contour is said to belong 
to the Keldysh space and in general such functions can be written as

\begin{equation}
k(z,z') =\delta(z,z')k^{\delta}(z) + \theta(z,z')k^>(z,z')+\theta(z',z)k^<(z,z')  
\end{equation}

where $\theta(z, z')$ is a contour Heaviside function, 
i.e., $\theta(z, z')=1$ for $z$ later with respect to the contour ordering than $z'$ on the
contour and zero otherwise, and $\delta(z, z' ) = \partial_z \theta(z, z') $ 
is the contour delta function. Examples of such quantities are of course the Green function, 
whose singular part $G^\delta$ is zero, and the self-energy, 
where the singular part is given by the Hartree-Fock self-energy $\Sigma^\delta = \Sigma^{HF} [G]$.
The greater and lesser term respectively denote the correlation parts. 
From this expression we can define several subordinated functions. 
It is common to denote by $z=t_-$ the contour point on the forward branch, $z=t_+$ 
the contour point on the backward branch and $z=t_0-i\tau$ the contour point on the vertical track. 
The Keldysh component lesser ($<$), greater ($>$), 
retarded ($R$), advanced ($A$), left ($\lceil$), right ($\rceil$) and Matsubara ($M$) 
of the function $k(z,z')$ on the contour are defined according to \cite{stef van}
\begin{align}
&k^M(\tau,\tau') = k(t_0-i \tau,t_0-i \tau')\\
&k^\lceil(\tau,t') = k(t_0-i \tau,t') \\
&k^\rceil(t,\tau) = k(t,t_0-i \tau) \\
&k^{\lessgtr}(t,t') = k(t_{\mp},t'_{\pm}) \\
&k^{R/A}(t,t') = \pm\theta(\pm(t-t'))[k^>(t,t')-k^<(t,t')]
\end{align}
With these definitions and with the help of Langreth rules for the projection
of integral of products of two Keldysh functions, 
we can write the equations for the real-time components of the Dyson equation:

\begin{equation}
\label{eq:dysonm}
G^{M} (\tau,\tau') = [G_0^{M} + G_0^{M} \star \Sigma^{M} \star G^{M}](\tau,\tau'),
\end{equation}

\begin{align}
\label{eq:dysonle}
G^{\lceil} (\tau,t') =& [G_0^{\lceil} + G_0^{\lceil} \circ \Sigma^{A} \circ G^{A} +  \\
&+G_0^{M} \star \Sigma^{\lceil} \circ G^{A} +G_0^{M} \circ \Sigma^{M} \star G^{\lceil}](\tau,t')\nonumber,
\end{align}

\begin{align}
\label{eq:dysonri}
G^{\rceil} (t,\tau) =& [G_0^{\rceil} + G_0^{R} \circ \Sigma^{R} \circ G^{\rceil} +  \\
&+G_0^{R} \circ \Sigma^{\rceil} \star G^{M} +G_0^{\rceil} \star \Sigma^{M} \star G^{M}](t,\tau)\nonumber,
\end{align}

\begin{equation}
\label{eq:dysonra}
G^{R/A} (t,t') = [G_0^{R/A} + G_0^{R/A} \circ \Sigma^{R/A} \circ G^{R/A}](t,t'),
\end{equation}

\begin{align}
\label{eq:dysonlg}
G^{\lessgtr} (t,t') &= [G_0^{\lessgtr} + G_0^{R} \circ \Sigma^{\lessgtr} \circ G^{A}  \\
&+ G_0^{\lessgtr} \circ \Sigma^{A} \circ G^{A} +G_0^{R} \circ \Sigma^{R} \circ G^{\lessgtr} \nonumber \\
&+ G_0^{\rceil} \star \Sigma^{M} \star G^{\lceil} \nonumber\\
&+ G_0^{R} \circ \Sigma^{\rceil} \star G^{\lceil}  + G_0^{\rceil} \star \Sigma^{\lceil} \circ G^{A} \nonumber](t,t').
\end{align}

The notations $\circ$ and $\star$ denote the real-time and imaginary-time convolutions:

\begin{align*}
&[a \circ b](t,t') = \int_{t_0}^\infty  a(t,\bar{t}) b(\bar{t}, t')  d \bar{t},   \\
&[a \star b](t,t') = -i \int_{t_0}^{t_0-i\beta}  a(t,\tau) b(\tau, t')  d \tau,
\end{align*}
where the matrix multiplication among indices of the single-particle basis is assumed.

Eq.~\ref{eq:dysonm} is the only equation which is completely decoupled from all other equations
and gives information on the initial state of the system at time $t_0$.
All other equations are coupled and their solution has to be found self-consistently as discussed in the 
next section.
Eqs.~\ref{eq:dysonle} and ~\ref{eq:dysonri} give the contribution of the initial state
to the dynamical properties of the system.
The retarded and advanced components given by the solution of Eqs.~\ref{eq:dysonra} give the spectral properties
of the system whereas the solution of Eqs.~\ref{eq:dysonlg} give the particle and hole propagators which in turn 
give access to occupancy numbers and time correlations in the interacting system.

\section{Self-energies}
In this section, we outline some of the most used self-energy approximation
which embody both the role of many-body interactions and of the external reservoirs.
Of course it is not at all an exhaustive list of them as the choice of the self-energy is strictly 
related to the system at hand. Our aim is only to give few examples in order to outline the main
features which can then be exploited in the numerical implementation.
On the other hand, the prototype self-energies considered here cover the most used ones for electronic
systems and to some extent also the only ones amenable of a numerical implementation with 
good scaling properties with both the size of the system and of the number of time steps.
This in turn can serve as a guide for other more specific and system-tailored approximation.

\subsection{Many-Body Self-energy}

We have seen that in order to include the effect of many-body interactions in the dynamics
of the single-particle Green's function it is possible to define a self-energy
which describes the effect of the whole system at the single particle level.

Formally the self-energy arise either as a way to truncate the the Martin-Swinger hierarchy 
or in the diagrammatic expansion of the evolution operator as a way to choose 
which physical processes are important in the description of the system.
In the first case one obtains the Kadanoff-Baym equations
whereas in the second one the Dyson equation or, 
in a more general framework, the set of Hedin equations.
In either case the choice of the self-energy is, to some extent, left to the needs of
the problem addressed, meaning that the choice of the diagrams to be included in the self-energy depends only upon the physical processes which are believed 
to contribute the most to the specific case at hand. 

On the other hand, there are some general restrictions to consider,
specifically those imposed by the macroscopic conservation laws.
To guarantee that the latter are preserved, such as conservation of particle, 
energy, and linear and angular momentum, 
the self-energy has to be the functional derivative of
a functional $\Phi[G]$ with respect to G, namely:

\begin{equation}
\Sigma (1;1') = \frac{\delta \Phi[G]}{\delta G(1;1')}.
\end{equation} 

In this case the resulting single-particle Green's function is guaranteed 
to fulfill macroscopic conservation laws.

This introduces a non-trivial problem if one attempts to find the interacting 
Green's function.
By definition the self-energy is a functional of the 
{\it interacting} single-particle Green's function which in turn can be found
only through the knowledge of the self-energy.
Therefore conservation laws
are satisfied if and only if a self-consistent procedure is employed.
This is one first obvious reason to resort to numerical techniques.

\begin{figure}[t]
\includegraphics[width=\linewidth]{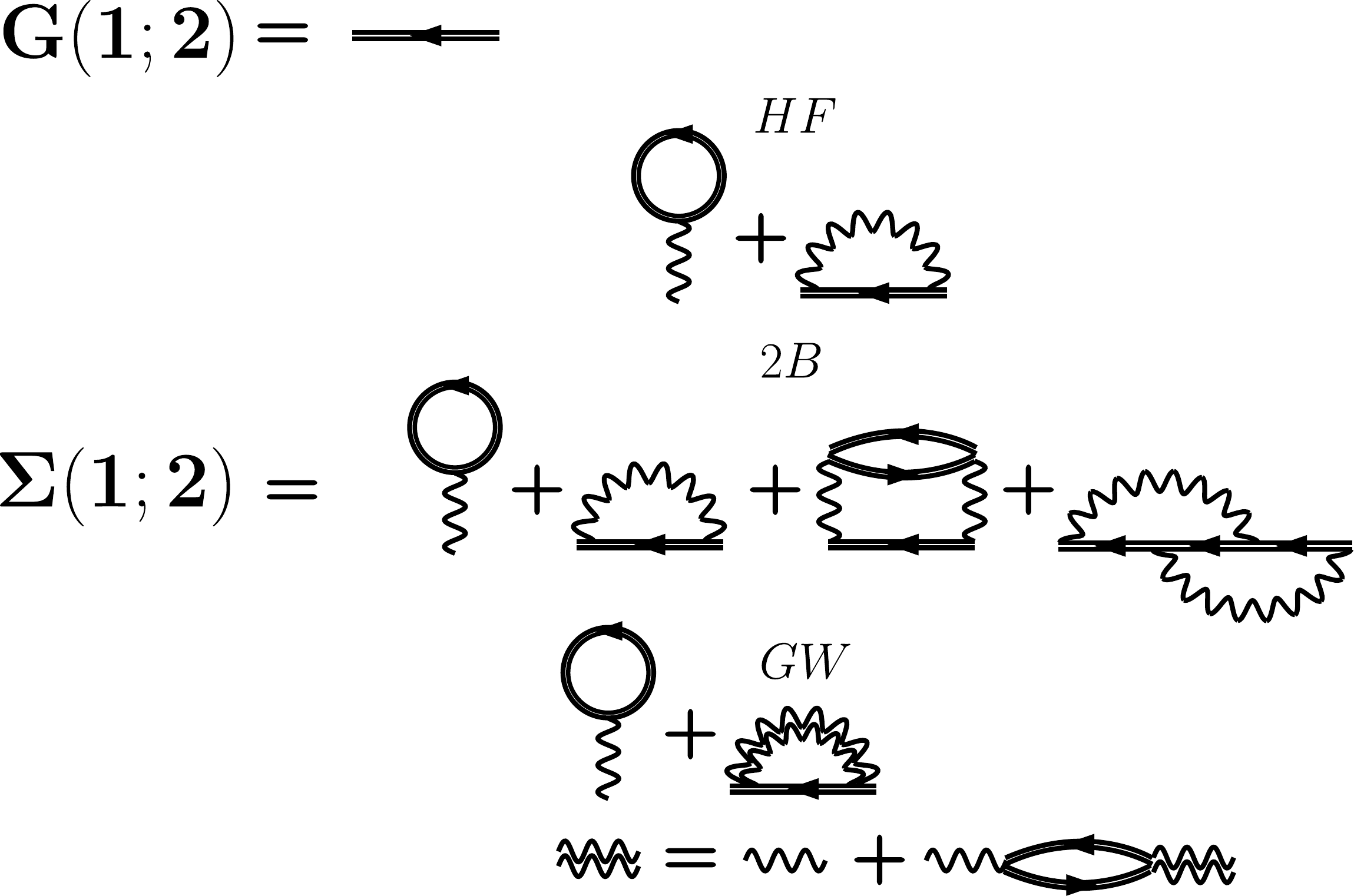}
\caption{Feynman diagrams constituting the Hartree-Fock (HF) second-Born (2B) ad GW (RPA) approximations
for the many-body self-energies.}
\label{fig2B}
\end{figure}

The most well-known conserving approximations are the Hartree-Fock (HF), the second Born (2B), 
the GW, and the T-matrix approximation. 
We present in the main text only the first three of these approximations. 
The first-order approximation for the self-energy, 
namely the HF approximation, has the following mathematical form

\begin{align}
\label{HF}
\Sigma^{HF} [G](1,1') &= -i \delta(1,1') \int d \bar{1} v(1, \bar{1}) G(\bar{1};\bar{1}^+) \nonumber \\
&+i G (1;1') v(1^+,1')
\end{align}

this approximation describes how a particle moves freely under the influence 
of an effective potential which depends of all the other particles, 
namely the HF approximation includes the effect of the interaction 
through an effective potential (mean-field approximation). 
Up to the second-order approximation for self-energy, 
the first example which one encounters is the 2B approx. that reads

\begin{align}
\label{secB}
&\Sigma^{2B} [G](1;1') = \Sigma^{HF} [G](1;1') \nonumber \\ 
&-i^2  G(1;1') \int d \bar{1} d\bar{2}v(1, \bar{1} )G(\bar{1}, \bar{2})  G( \bar{2},\bar{1}) v(1',\bar{2}) \nonumber \\
&+i^2 \int d \bar{1} d\bar{2} v(1, \bar{1} )G(1;\bar{2})G(\bar{2}, \bar{1})  G( \bar{2},1') v(1',\bar{2})  
\end{align}

here, in addition to the time-local part of the self-energy ($\Sigma^{HF}$), 
we have terms up to the second order in the Coulomb interaction $v(z,z')= v(\bold{x}, \bold{x'}) \delta(t,t')$. 
The first term after the HF-part of the self-energy is generally denoted 
as first order bubble diagram Fig.~\ref{fig2B} which describes a propagation of a particle (or hole) 
while interacting with particle hole-pair, i.e., 
it includes to first order the polarization of the media due to inserted particles (or hole). 
The last term is nothing but the second order correction to the exchange term.  \\

In the $GW$ approximation the electronic self-energy takes the form 

\begin{equation}
\Sigma^{GW}[G] (1;1') = \Sigma^{H}(1,1') + i W(1;1')G(1;1') 
\end{equation}
 
with $\Sigma^{H}$ being the Hartree part of the self-energy, the first term of the r.h.s of Eq. \eqref{HF}, and where the dynamically screened interaction $W$ satisfy the Dyson equation 

\begin{equation}
W(1;1') = v(1,1') + \int d \bar{1} d \bar{2}  v(1,\bar{1}) P(\bar{1},\bar{2}) W(\bar{2}, 1'),
\end{equation}
the polarization is usually approximated in the random-phase approximation (RPA) as $P(z,z') = -i G(z,z')G(z',z)$.
The $GW$ approximation can be seen as a dynamically screened exchange 
approximation able to describe the effects of long-range interaction. 

Other choices for the polarization diagram are possible~\cite{stef van}, but unfortunately they typically make the computation more
demanding as a vertex function has to be included.
In fact this case corresponds to including in the set of equation to 
be solved a vertex functional which in turn means to switch to a more complicated set of equations then the ones including only the Green's function and the self-energy. 
This set, named Hedin equations, is made of five equations 
which have to be solved at the same time and self-consistently.
The complication arises mostly due to the nature of the vertex functional 
which is, in general, a three indexes object and has a more complicated 
structure in Keldysh space than the functions seen so far.

\subsection{Embedding self-energy}
Moreover, the self-energy in \eqref{dysongreen} should take into account the tunneling of electrons between the central region and the leads. We define the embedding self-energy to be proportional to the non-interacting lead Green's function in the localized (dot) basis and to the coupling hamiltonian as  

\begin{equation}
\label{embself}
\Sigma_{em,\alpha, kl} (z,z') = \sum_{ij}^{\mathcal{N}_\alpha} T_{k, i \alpha} g_{\alpha \alpha, ij} (z,z') T_{ j \alpha, l}  
\end{equation} 

here $k,l$ label the sites in the central region, while $i,j$ label the sites in the lead $\alpha$. Furthermore, $\mathcal{N}_\alpha$ is the number of sites in the lead $\alpha$ (at the end of the derivation we take $\mathcal{N}_\alpha \to \infty$). 
The non-interacting Green's function in delocalized basis (in the eigenbasis of the leads with eigenvalues $\epsilon_{i \alpha}$) can be readily found to be 

\begin{align}
\tilde{g}_{\alpha \alpha, ij}^< (t,t') &= i \delta_{ij} f(\epsilon_{i \alpha}) e^{-i \int_{t'}^{t}(\epsilon_{i \alpha} -\mu +V(\bar{t})) d \bar{t}} \label{gllead} \\
\tilde{g}_{\alpha \alpha, ij}^> (t,t') &= i \delta_{ij} (f(\epsilon_{i \alpha})-1) e^{-i \int_{t'}^{t}(\epsilon_{i \alpha} -\mu +V(\bar{t})) d \bar{t}} \label{gglead} 
\end{align}

where $\mu$ is the chemical potential and $V_\alpha(t)$ is the bias voltage in the lead $\alpha$, while the function $f(\epsilon)= 1/ (e^{\beta (\epsilon -	\mu)}+1)$ denotes the Fermi distribution function. We can express the lead Green's function in the localized site basis via a basis transformation which diagonalizes the lead Hamiltonian $D^{\dagger} \hat{H_\alpha} D =$ diag$[\epsilon_{i \alpha}]$ such that 

\begin{equation}
g_{\alpha \alpha, ij}^{\lessgtr}(t,t') = \sum_{kl}^{\mathcal{N}_\alpha} D_{ik}^\alpha \tilde{g}_{\alpha \alpha, kl}^{\lessgtr} (t,t') D_{lj}^{\alpha \dagger} .
\end{equation} 

Inserting Eqs. \eqref{gllead} and \eqref{gglead} into the definition of the embending self-energy \eqref{embself} and defining the function 

\begin{equation}
\Gamma_{kl, \alpha} (\epsilon) = 2 \pi   \sum_{ij n}^{\mathcal{N}_\alpha} T_{k, i \alpha} D_{in}^\alpha \delta(\epsilon-\epsilon_{n \alpha})  D_{nj}^{\alpha \dagger} T_{ j \alpha, l}
\end{equation}

we get an expression for the one-dimensional lead self-energy

\begin{align}
\Sigma_{em,\alpha, kl}^{<} (t,t') &= i e^{-i \int_{t'}^{t} V_\alpha(\bar{t}) d \bar{t}} \times \nonumber \\
&\times \int \frac{d \epsilon}{2 \pi} f(\epsilon) \Gamma_{kl, \alpha} (\epsilon) e^{-i(\epsilon-\mu)(t-t') } 
\end{align}

\begin{align}
\Sigma_{em,\alpha, kl}^{>} (t,t') &= i e^{-i \int_{t'}^{t} V_\alpha(\bar{t}) d \bar{t}} \times  \nonumber \\
&\times \int \frac{d \epsilon}{2 \pi} (f(\epsilon) -1)\Gamma_{kl, \alpha} (\epsilon) e^{-i(\epsilon-\mu)(t-t') } 
\end{align}

Furthermore, the $\lceil, \rceil$ and $M$ component are similarly worked-out by considering the time-arguments on different parts of the Keldysh contour.

\subsection{Symmetries}

Before moving to the next section, in which we will present the numerical schemes used to solve the Dyson equation it is useful to recall the symmetry properties, in the  single-particle basis of the system, between the different real and imaginary time axis of both Green's functions and self-energies which can be exploited in order to 
reduce the computational load.
They are given (for fermions) by:

\begin{align}
&G_{ji}^{M} (\tau;\tau') =  -G_{ij}^{M} (\tau;\tau')^*\\  
&G_{ij}^{\lceil} (\tau;t) =  G_{ji}^{\rceil} (t;\beta-\tau)^* \\
&G_{ji}^{A} (t';t) =  G_{ij}^{R} (t;t')^*  \\
&G_{ji}^{\lessgtr} (t';t) = - G_{ij}^{\lessgtr} (t;t')^*
\end{align}

\begin{align}
&\Sigma_{ij}^{M} (\tau;\tau') = - \Sigma_{ji}^{M} (\tau;\tau')^* \\
&\Sigma_{ij}^{\lceil} (\tau;t) =  \mp \Sigma_{ji}^{\rceil} (t;\beta-\tau)^* \\
&\Sigma_{ji}^{A} (t';t) =  \Sigma_{ij}^{R} (t;t')^*  \\
&\Sigma_{ji}^{\lessgtr} (t';t) = - \Sigma_{ij}^{\lessgtr} (t;t')^* 
\end{align}

The importance of the symmetries in this context is that they allow 
to reduce both the computational time and, also very important, the memory required.

\section{Physical Quantities}

In this section we present some physical quantities which can be calculated through the knowledge of the 
interacting single-particle Green's functions and the self-energy.
Specifically, we discuss quantities which give information on the dynamics of the system
and which therefore require the knowledge of the the two-time Green's function.

\subsection{Occupation number and momentum distribution}
The first physical quantity of interest is the occupation number of a given single particle state,
which can be computed by means of the lesser Green's function as:

\begin{equation}
n_s(t) = -iG^{<}_{s,s}(t;t),
\end{equation}
this quantity gives us information on the time-dependent occupation of the state with quantum numbers $s$.

In the case in which the chosen basis is made of sites and spin, namley $s=\{{\bf j},\sigma\}$ where ${\bf j}$ labels the 
position of a site ${\bf j}$ in a lattice of N sites in d-dimensions, then one can define the momentum distribution.
In the case

\begin{equation}
m({\bf k}) = -i\frac{1}{N}\sum\limits_{\sigma}\sum\limits_{\sigma}e^{-i({\bf j}_1-{\bf j}_2)\cdot{\bf k}}G^{<}_{j_1\sigma,j_2\sigma}(t;t).
\end{equation}
This case is particularly relevant in the case of ultracold gases where the momentum
distribution is a routinely measured quantity via time-of-flight techniques.

\subsection{Spectral Function} 
Information on the single particle spectrum are encoded in the spectral function, 
defined as the Fourier transform of $G^> -G^<$ with respect to the relative time coordinate $\tau = t-t'$ 
for a given value of the time $T=(t+t')/2$, i.e.,

\begin{equation}
A(T, \omega) = -\text{Im} \int \frac{d \tau }{2 \pi} e^{i \omega \tau } \Omega \left(T+\frac{\tau}{2},  T-\frac{\tau}{2} \right). 
\end{equation}
where $\Omega=  \text{Tr}_{\bar{C}}\left[G^> - G^< \right]$ and the trace is over all degrees of freedom not belonging to the central region (system of interest).
For values of $T$ sufficiently large, where the transient dynamics becomes negligible, and in the absence of external drivings,
the spectral function becomes independent of $T$, namely $A(T,\omega)\rightarrow A(\omega)$. 
For these times $\text{Tr}_C A(\omega)$ displays peaks at the accessible single-particle energies, namely it is propostional to the non-equilibrium density of states (DOS).
The spectral function also contains other informations such as the momentum distribution of the system and therefore to the single-particle
dispersion relation thanks to the information on the frequencies $\omega$.

It is worth stressing that the calculation of the spectral function requires the knowledge of the two-time Green's function
and therefore the information it carries is not accessible in the context of the GKBA.
Actually in this last case this information has to be provided {\it a priori} via the ansatz on the Retarded and Advanced 
components of the Green's function and because so far the most used approximation used is the Hartree-Fock, this amounts
to use a single-particle spectrum calculated at the HF level and therefore lacking of any many-body correlation effect.

\subsection{Currents}
From the knowledge of the single particle Green's function of the central region Eq. \eqref{green} and its components along the Keldysh contour we have access to the time-dependent ensemble average of all one-body operators of the central region. We also have access to the total current flowing between this region and the lead $ \alpha$. 
The particle current reads

\begin{equation}
I_\alpha (t) = - \left\langle  \frac{d}{dt} \hat{N} (t) \right\rangle = -i \left\langle \left[ \hat{H}_T^\alpha (t), \hat{N} (t) \right]  \right\rangle
\end{equation}
where $\hat{N} (t)$ is operator related to the total number of particles in the system whereas $\hat{H}_T^\alpha (t)$
is the coupling Hamiltonian between the central region and the leads $\alpha$.
and one can easily rearrange it in terms of the single particle Green's function of the central region 
and the embedding self-energy of the lead $\alpha$ by means of the equation of motion~\cite{stef van}:

\begin{equation}
\label{current}
I_\alpha (t) = 2 Re \left \{  Tr \left[ \Sigma_\alpha^< \cdot G^A + \Sigma_\alpha^R \cdot G^< + \Sigma_\alpha^\rceil \star G^\lceil  \right] (t,t) \right \}
\end{equation}

The last term in \eqref{current} explicitly accounts for the effects of initial correlations and initial-state dependence. If one assumes that both dependencies are washed out in the limit $t \to \infty$, then for large times we can discard the imaginary time convolution. 
The resulting formula is known as the Meir–Wingreen formula. 
Equation \eqref{current} provides a generalization
of the Meir–Wingreen formula to the transient time-domain.

Another important current is the one given by an expression completely analogous to the one in Eq.~\ref{current}
by replacing the embedding self-energy with the many-body one. In the following we will label this current $I_{MB}(t)$.
This current comes from the interaction term $\hat{V}(t)$ and therefore has to go to zero due to the fact 
that this term conserves the total number of particles because it commutes with the central region's Hamiltonian 
at any time.
On the other hand, as we discussed above, this is only true if self-consistency holds;
therefore this quantity is a good figure of merit of the convergence of the self-consistence procedure.
We will show an example below.

\section{Numerical implementation and self-consistent solution of the Dyson equation}

\begin{figure}[t]
\begin{center}
\includegraphics[width=.5\textwidth]{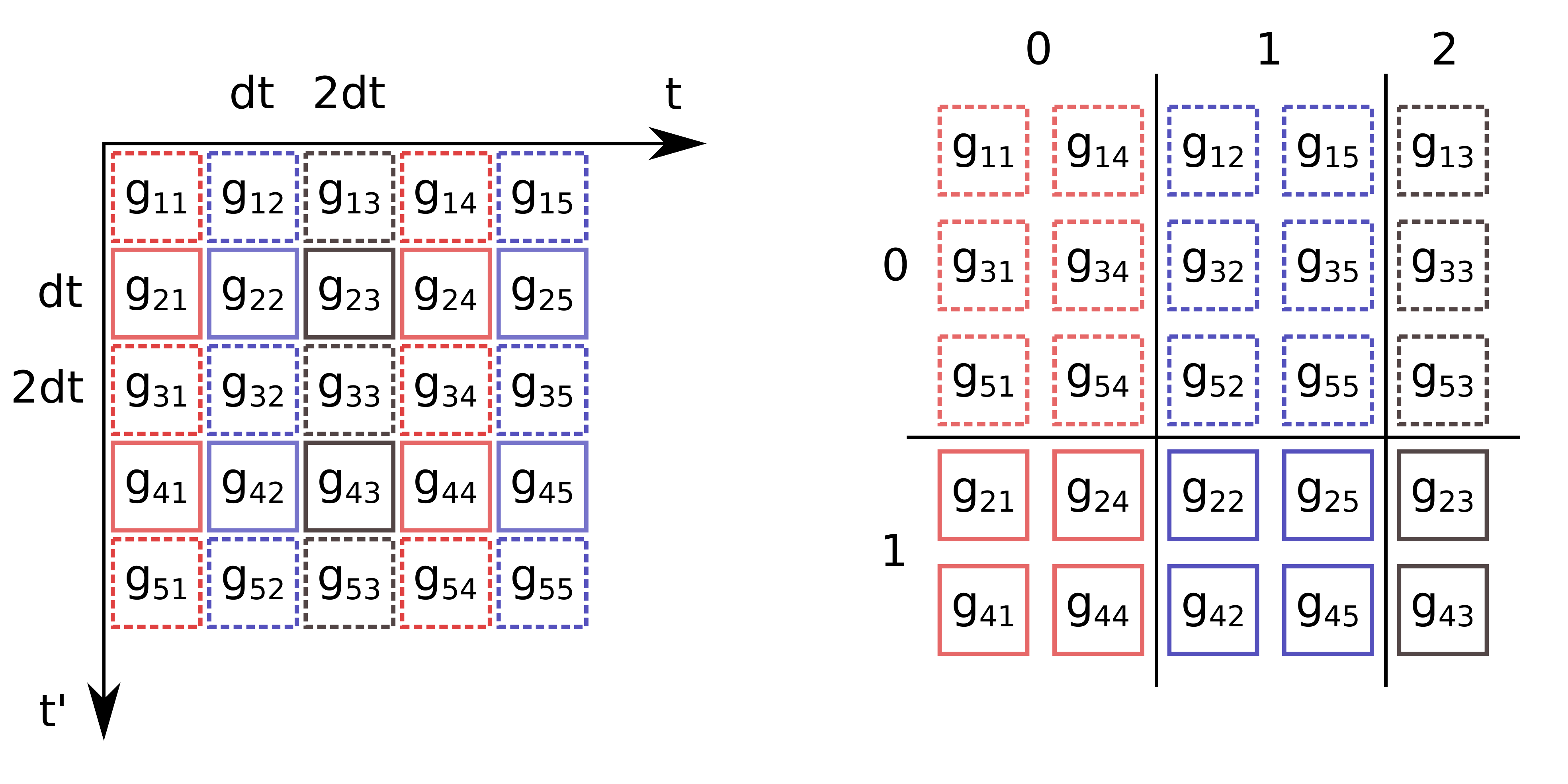}
\caption{Distribution of the spatial blocks $g_{ij}$ of dimension $n_s \times n_s$ of the full matrix $n_s n_t \times n_s n_t$ according to the $2$-dimensional block-cyclic distribution scheme (2DBCD) into a process grid $P_{r} \times P_{c}$ generated by using a row-major ordering of the processes (RMOP). The example refers to a full matrix of dimension $n_t \times n_t = 5\times 5$ with $n_s=1$ and $g_{ij}$ being just its matrix elements into a process grid of $6$ processes }
\label{figPD}
\end{center}
\end{figure}

In order to solve the Dyson Eqs.~\eqref{eq:dysonm}-\eqref{eq:dysonlg} we store each function into a matrix 
$(n_s n_t) \times (n_s n_t)$ where $n_s$ is the 
number of spatial and spin (or momentum and spin) points
and $n_t$ the number of points in which the time interval
($[t_0,t_f]$ for the real axis and $[t_0-i\beta,t_0]$ for the imaginary one) considered is divided into.
The solution of the the Dyson Eqs.~\eqref{eq:dysonm}-\eqref{eq:dysonlg} requires two matrix inversions, one for the Matsubara Green's function and the other for the Retarded one, and several matrix multiplication.
Furthermore one extra inversion is required in the case of the $GW$ self-energy to solve the Dyson-like equation for the dressed interaction.
If performed with a standard sequential algorithm, 
these operation would require a time of the order $O(N^3)$ with N the dimension of the matrix. 
Even if more refined schemes would be used, the scaling 
of the computational time with the 
matrix size would be of the order $O(N^\alpha)$ with $2<\alpha<3$.
The structure of the equations suggests one obvious choice to improve the scalability of the computational time with the size of the matrices, namely the recourse to parallel schemes to perform matrix operations. 
To exploit parallel algorithms for matrix operation we distribute each matrix on a grid of $p=p_r\times p_c$ processes. With such a distribution we are able to exploit
the Scalable Linear Algebra Package (ScaLAPACK) routines which employ Message Passage Interface (MPI) protocol through the Basic Linear Algebra Communication Subprograms (BLACS)
routines to handle communication between processes and perform matrix operations.
This choice relies on the fact that the ScaLAPACK routines 
together with the BLACS offer a user-friendly interface 
to exploit the MPI potentiality together with the 
advantage of employing well-tested and optimized routines 
to perform matrix operations and communication tasks.
After initialization of a $p_{r} \times p_{c}$ process grid by using a row-major ordering of the processes (RMOP)
we distribute the blocks of each matrix onto the process grid via a $2$-dimensional block-cyclic distribution scheme (2DBCD) as shown in Fig.~\ref{figPD}
where we illustrate an example of distribution of a matrix $g$ with size $n_s n_t \times n_s n_t$  for $n_t=5$ ; 
the elements of $g$ are distributed on $6$ processes organized into a process grid  $p_{r} \times p_{c}=2 \times 3$.

One key aspect in handling any parallelized algorithm is to minimize communication 
between processes in such a way to avoid communication bottle necks which would spoil 
any advantage of resorting to parallel computation.  
In our case particular care has to be taken in the calculation of the self-energies, which is usually 
the critical point of algorithms based on many-body perturbation theory.
A key observation is that all the self-energies discussed above are bi-local in both time arguments. 
On the other hand, sums over spatial (momentum) indexes are needed. Therefore we have chosen to distribute the matrices in such a way that each process stores chunks 
of the whole matrix which corresponds to the full spatial (or momentum) indexes for a given pair of times $(t,t')$, $(t,\tau)$, $(\tau,t')$, $(\tau,\tau')$.
In this way the computation and the assignment of the different chunks of the self-energies can be performed locally on each process and require no communication.
The only exception among the discussed self-energy approximations is given by the 
$GW$ self-energy where the integral equation for the dressed interaction is solved 
via a matrix inversion handled by the ScaLAPACK routines.
Another example which is worth mentioning is that of the T-matrix
approximation where an extra inversion is required 
in order to solve the Bethe-Salpeter equation 
for the two-particle scattering amplitude~\cite{LoGullo2016}.

The Dyson Eqs.~\eqref{eq:dysonm}- \eqref{eq:dysonlg}
are Volterra integral equation of second kind are mapped into a linear system of algebraic equations after discretization of the time variable.
This allows to exploit different integration schemes to reduce the error in the time step 
analogously to what can be done in the case of the
GKBs by choosing higher order integration schemes~\cite{numericalrecipes}.
The self-consistent solution is then found, primarily, by solving the equation for the Matsubara Green's function
starting from $G_{0}^M (\tau;\tau')$ which is used to compute the Matsubara self-energy $\Sigma^M (\tau;\tau')$
and then iterating the procedure until convergence.
Once the Matsubara components of the interacting Green's function and self-energy are found, 
all other equations are solved similarly by starting 
from the solution $G_{ij}^{(0)} (1;1')$ of the non-interacting equation  \eqref{nnintgreen}
which is used to compute the self-energies all other self-energies.
Once again this procedure continues until convergence.

\section{Quantum quench in fermionic ultracold gases}
In this section we use the technique described above to study the quench of an ultracold gas 
in an optical lattice. Specifically, we look at the dynamics of a one-dimensional 
system of fermionic atoms with spin-$1/2$ described by the Fermi-Hubbard model:

\begin{align}
 \hat{H}&=\hat{H}_0+\hat{V}\\
 \hat{H}_0&=\sum\limits_{n \sigma} \epsilon_{n}\hat{c}_{n \sigma}^{\dag}\hat{c}_{n \sigma}
 -\frac{J}{2}\left(\hat{c}_{n+1 \sigma}^{\dag}\hat{c}_{n \sigma}+\text{h.c.}\right)\\
 \hat{V}&=\frac{U}{2}\hat{n}_{n \uparrow}\hat{n}_{n \downarrow},
 \label{eq:ham}
\end{align}

where $\epsilon_{n}=\lambda\cos(2\pi\tau n)$ ($\tau=\left(\sqrt{5}+1\right)/2$)
are the on-site energies, $U$ the on-site interaction between particles with different spin in the s-wave approximation, $\hat{c}_{n \sigma}^{\dag}(\hat{c}_{n \sigma})$ are fermionic creation (annihilation) operators at site $i$ with spin $\sigma$ and $\hat{n}_{n \sigma}=\hat{c}_{n \sigma}^{\dag}\hat{c}_{n \sigma}$ the corresponding number operator. Eq.~\ref{eq:ham} defines the interacting Aubry-Andr\'e model.
An interesting feature of this model is that for $U=0$ it witnesses a metal-to-insulator transition at $\lambda=1$,
specifically for $\lambda<1$ the eigenstates are all delocalized whereas for $\lambda>1$ all eigenstates are exponentially localized in the thermodynamic limit~\cite{Jitomirskaya1999}.

We consider an initial state in which even sites are occupied by two particles, one per spin degree of freedom,
with no correlations between the two spin degrees of freedom. Odd sites are empty.
This state can be realized routinely in these type of experiments by increasing the height of the optical lattice in 
order to forbid tunneling of particles to other sites and tuning the interactions between particles by means of Feschbach resonances.
At time $t_0=0$ the barriers between the different sites are lowered and interaction is increased to the desired value. This procedure defines the quench protocol.
As a consequence of the quench the system is brought out-of-equilibrium and the post-quench dynamics is
what we want to study here.

\begin{figure}[t]
\begin{center}
\includegraphics[width=.5\textwidth]{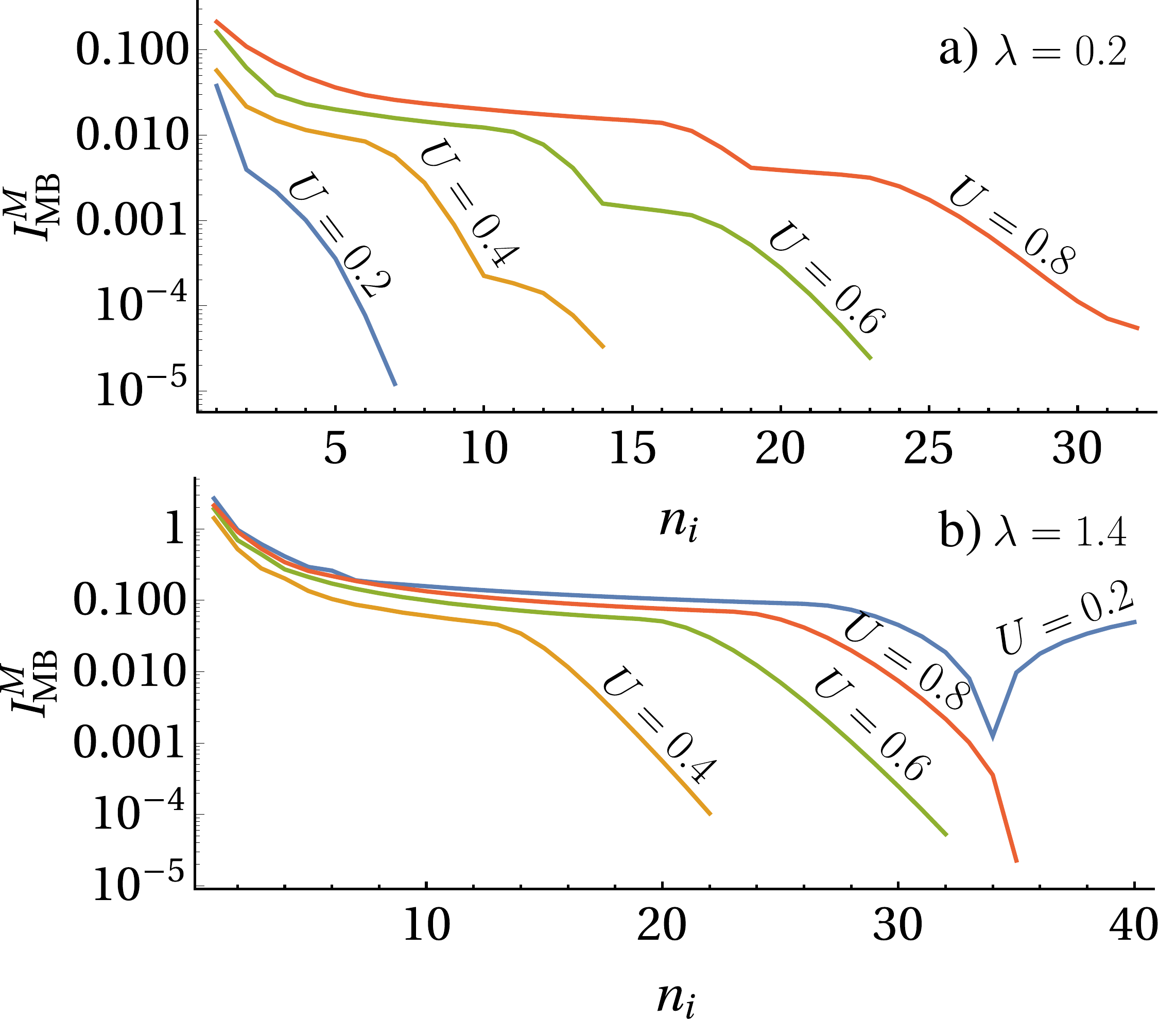}
\caption{Maximum value of the many-body current $I_{MB}(t)$ as a function of the number of iterations $n_i$ 
of the self-consistent scheme for different values of $\lambda$ and $U$.
Panel a) $\lambda=0.2$ (below the non-interacting transition point) and panel b) $\lambda=1.4$ (above the non-interacting transition point).}
\label{fig:conv}
\end{center}
\end{figure}

For this specific problem we do not need neither the propagation of the Matsubara Green's function, nor that
for the left or right ones because the initial state is a non-interacting one given by a Gibbs
state of which the density matrix operator has the form $\hat{\rho}=e^{-\beta(\hat{H}_0-\mu\hat{N})}/Z$ where $\beta$ is the inverse temperature $(k_B=1)$,
$\mu$ the initial chemical potential, $\hat{N}=\sum\limits_{n \sigma} \hat{c}_{n \sigma}^{\dag}\hat{c}_{n \sigma}$
the total number of particles and $Z=\text{Tr} e^{-\beta(\hat{H}_0-\mu\hat{N})}$ the partition function.
Therefore we are left with only three equations to be solved,
namely that for the Retarded, Lesser and Greater Green's functions.
We will study the dynamics of the system for different interactions $U$ and height of the optical potential $\lambda$
for a system with $n_s=40$ sites and with initially $N=40$ particles equally distributed between over the two degrees of freedom.
Both the number of time points $n_t$ and also the final time $t_f$ are chosen in the range $[900-980]$ and $[40-60]$
depending on both $\lambda$ and $U$. The final time of evolution has been chosen such that the quantity $\delta \omega =1/t_f$
is of the order of the smallest energy level spacing of the spectrum of $\hat{H}_0$ and consequently the number of time points
is such that $1/dt$ with $dt=t_f/(n_t-1)$ is at least two times larger than the larger eigenvalue of $\hat{H}_0$ (we are taking the 
spectrum of $\hat{H}_0$ centered around zero).
This empirical estimation guarantees that we can resolve the dynamics of all the eigenvalues of the non-interacting Hamiltonian
and we are able to capture the coherences between different non-interacting eigenstates.

Due to the short range nature of the interaction in this 
system we use the second-Born self-energy approximation.

\subsection{Convergence of the scheme}

Before discussing some physical quantities of the system it is worth looking at the convergence properties
of the numerical scheme presented and discussed in the previous sections.

In Fig.~\ref{fig:conv} we show the maximum over all time steps of the many-body current $I_{MB}^M= \text{MAX}_t[|I_{MB}(t)|]$.
As we discussed above, the many-body current has to go to zero because it formally arises from the many-body self-energy 
and its vanishing is a signature of the self-consistent nature of the scheme we are using.

We can see that in the (non-interacting) metallic ($\lambda=0.2$) as we increase the interaction more steps are required to reach convergence,
a result which has to be expected and can be explained with a qualitative argument. By increasing the interaction, 
the contribution of higher order diagrams counts more and therefore more iteration steps are required.
In the insulating phase $\lambda=1.4$ we observe an interesting feature, namely for small interactions ($U=0.2$ in figure) 
the convergence is really slow and at one point the many-body current starts to increase again. 
This is of course due to the fact that we are in a 
localized phase even in the presence of interaction and therefore the perturbation theory could diverge due to numerical instabilities.
On the other hand as we increase the interactions, the convergence of the scheme is recovered thus confirming that the 
anomalous behavior for small interactions is indeed due to the localization of the system and not to the value of the interaction itself.

\begin{figure}[!t]
\begin{center}
\includegraphics[width=\linewidth]{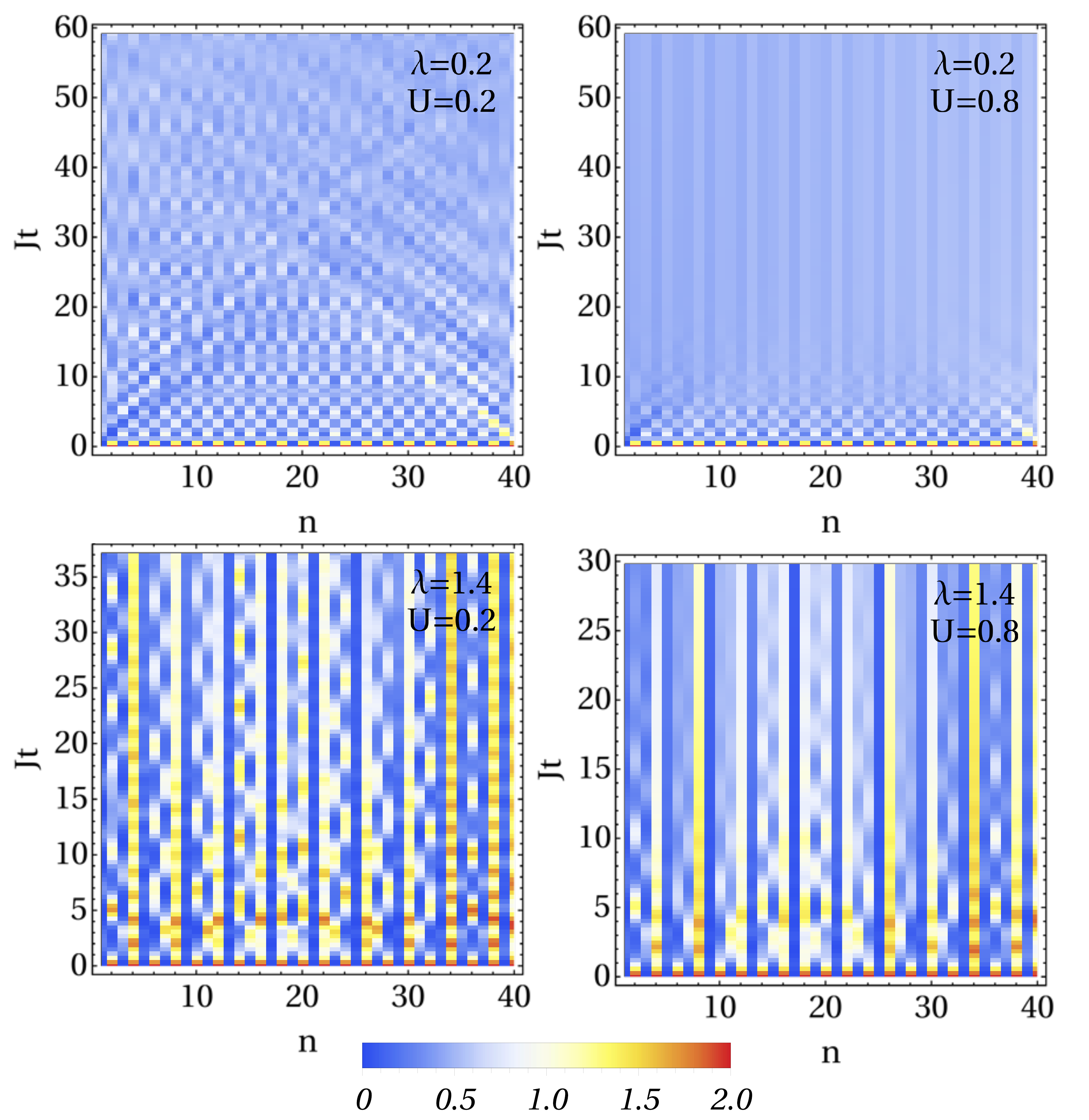}
\caption{Density of particles as a function of the site $n$ 
and time following the sudden quench for different values of $\lambda$ and $U$.}
\label{fig:density}
\end{center}
\end{figure}

\subsection{Dynamical quantities and spectral function}
In order to study the dynamics of the system we look at the change
in the local density of particles as a function of time.
The results are shown in Fig.~\ref{fig:density}.
We can see that in the metallic phase interferences are present in the dynamics which nevertheless quickly die out 
as the interaction is increased.
On the other hand, in the insulating phase, the effect of the interaction is to restore the interferences which are lost
at small interactions due to localization effects.

\begin{figure}[!t]
\begin{center}
\includegraphics[width=\linewidth]{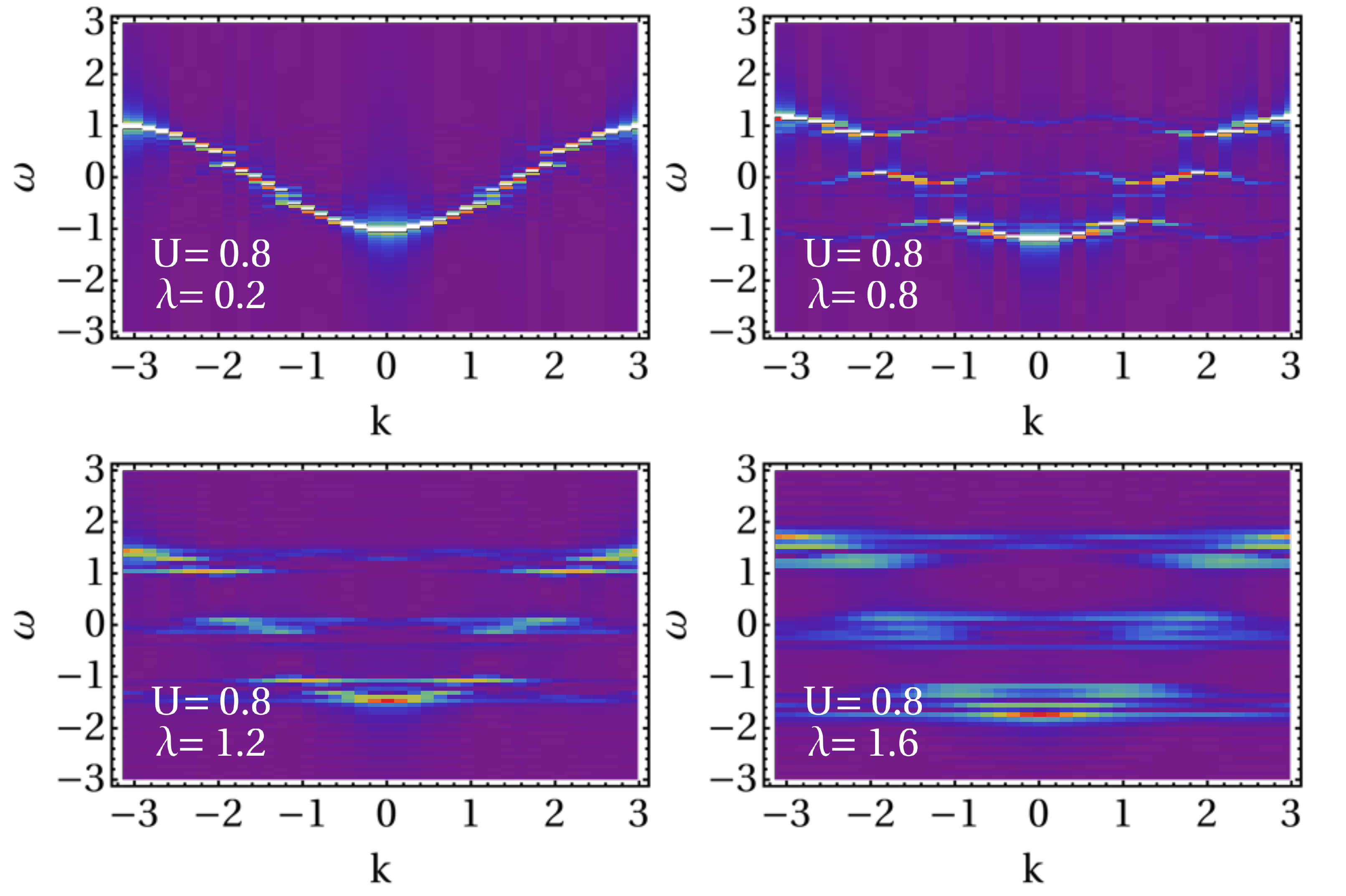}
\caption{Spectral function at long times for different values of $\lambda$ and $U$.}
\label{fig:specfun}
\end{center}
\end{figure}

In Fig.~\ref{fig:specfun} we show the spectral function 
$A(t_f,\omega)$ for different values of $\lambda$ at fixed interaction $U$.
It is interesting to notice that the localization properties 
just discussed are clearly visible in the broadening in momentum.
The effect of increasing $\lambda$ is to widen the gaps
in the system.

\section{Conclusion}
In this work we presented a scalable numerical approach to the solution of the Dyson-equation 
for the single-particle Green's function of a system of fermions mutually interacting and possibly coupled
to external reservoirs to study transport in correlated systems.
After reviewing the key ingredients of the non-equilibrium Green's function theory, we have introduced the main equations
to be solved in order to find the interacting single-particle Green's function.
Unlike in the standard propagation methods of the Kadanoff-Baym equations, we propose to solve the Dyson equation
which is of course completely equivalent to the former.
In both cases a self-energy functional has to be introduced to account for time-correlations arising from interactions,
interactions in the initial state and the propagation of the initial correlations in the dynamics of the system.
The numerical scheme is based on the mapping of the full Green's functions onto matrices with a double index structure,
one for the quantum number and the other for the time parameter.
In this form, the matrices can be distributed onto a grid of processes to lower the computational cost 
by exploiting parallel algorithms to perform matrix operations.
A key ingredient is to realize that the computation of the most used self-energy approximations
can be performed locally on each process by a clever choice of the distribution of the matrix onto the process grid.
In the end we show how our technique works by looking at the case of a quantum quenches in a system of ultracold fermionic atoms
with spin-1/2. In this example we discuss the convergence properties of the scheme we used and some interesting physics
in the dynamics of the gas.

\section{Acknowledgments}
The authors acknowledge financial support from the Academy of Finland Center of Excellence
program (Project no. 312058) and the Academy of Finland
(Project no. 287750).
NLG acknowledges financial support from the Turku Collegium for Science and Medicine (TCSM).
Numerical simulation were performed exploiting the Finnish CSC facilities under the Project no. 2001004 
("Quenches in weakly interacting ultracold atomic gases with non-trivial geometries").


\begin{thebibliography}{[1]}
\bibitem{greiner} M. Greiner, O. Mandel, T. Esslinger, T.W. Hänsch and I. Bloch,  Nature 415 39–44, (2002).
\bibitem{arijeet} Arijeet Pal and David A. Huse, Phys. Rev. B 82, 174411  (2010).
\bibitem{bardarson} Jens H. Bardarson, Frank Pollmann, and Joel E. Moore, Phys. Rev. Lett. 109, 017202 (2012).
\bibitem{serbyn} Maksym Serbyn, Z. Papić, and Dmitry A. Abanin, Phys. Rev. Lett. 111, 127201 (2013).
\bibitem{fausti} D. Fausti et al., Science 331, 189 (2011). 
\bibitem{kandelaki} Ervand Kandelaki and Mark S. Rudner,Phys. Rev. Lett. 121, 036801 (2018).
\bibitem{tddmrg} F. Heidrich-Meisner, A. E. Feiguin, and E. Dagotto, Phys. Rev.B 79, 235336 (2009).
\bibitem{nrg1} W. Izumida, O. Sakai, and S. Suzuki, J. Phys. Soc. Jpn. 70, 1045 (2001).
\bibitem{nrg2} W. Izumida and O. Sakai, J. Phys. Soc. Jpn. 74, 103 (2005).
\bibitem{frg} C. Karrasch, V. Meden, and K. Sch\"{o}nhammer, Phys. Rev. B 82, 125114 (2010).
\bibitem{frg1} J. Eckel, F. Heidrich-Meisner, S. G. Jakobs, M. Thorwart, M. Pletyukhov, and R. Egger, New J. Phys. 12, 043042 (2010).
\bibitem{frg2} S. G. Jakobs, M. Pletyukhov, and H. Schoeller, Phys.Rev.B 81, 195109 (2010).
\bibitem{diagram} K. S. Thygesen and A. Rubio, J. Chem. Phys. 126, 091101 (2007).
\bibitem{diagram1} A.-M. Uimonen, E. Khosravi, A. Stan, G. Stefanucci, S. Kurth, R. van Leeuwen, and E. K. U. Gross, Phys. Rev. B 84, 115103 (2011).
\bibitem{qmc} P. Werner, T. Oka, and A. J. Millis, Phys. Rev. B 79, 035320 (2009).
\bibitem{qmc1} P. Werner, T. Oka, M. Eckstein, and A. J. Millis, Phys. Rev. B 81, 035108 (2010).
\bibitem{Karlsson2018} D. Karlsson, R. van Leeuwen, E. Perfetto, G. Stefanucci, arXiv:1806.05639v1 (2018).
\bibitem{LoGullo2016} N. Lo Gullo, and L. Dell'Anna. Phys. Rev. B 94, 184308 (2016).
\bibitem{dft} K. Scheonhammer, O. Gunnarsson,  and R. M. Noack, Phys. Rev. B 52, 2504 (1995).
\bibitem{dft1} N. A. Lima, M. F. Silva, L. N. Oliveira, and K. Capelle, Phys. Rev. Lett. 90, 146402 (2003).
\bibitem{dft2} F. Maletand P. Gori-Giorgi, Phys. Rev. Lett. 109, 246402 (2012).
\bibitem{dft3} J.  Lorenzana,   Z.-J.  Ying, and  V.  Brosco, Phys. Rev. B 86, 075131 (2012).
\bibitem{dft4} A.  Mirtschink,  M.  Seidl, and  P.  Gori-Giorgi, Phys. Rev. Lett. 111, 126402 (2013).
\bibitem{tddft} C. Verdozzi, Phys. Rev. Lett. 101, 166401 (2008).
\bibitem{tddft1} S. Kurth, G. Stefanucci, E. Khosravi, C. Verdozzi, and E. K. U. Gross, Phys. Rev. Lett. 104, 236801 (2010).
\bibitem{tddft2} A.-M. Uimonen, E. Khosravi, A. Stan, G. Stefanucci, S. Kurth, R. van Leeuwen,   and E. K. U. Gross, Phys. Rev. B 84, 115103 (2011).
\bibitem{tddft3} A. Kartsev, D. Karlsson, A. Privitera, and C. Verdozzi, Sci. Rep. 3, 2570 (2013).
\bibitem{tddft4} M.  J.  P.  Hodgson,  J.  D.  Ramsden,  J.  B.  J. Chapman,  P.  Lillystone, and  R.  W.  Godby, Phys. Rev. B 88, 241102 (2013).
\bibitem{tddft5} M. J. P. Hodgson, J. D. Ramsden,  and R. W. Godby, Phys. Rev. B 93, 155146 (2016).
\bibitem{kadanoff}  L. P. Kadanoff, G. A. Baym,  Quantum statistical mechanics: Green’s function methods in equilibrium and non-equilibrium problems; Benjamin, (1962).
\bibitem{keldysh} L. V. Keldysh, Diagrammatic technique for non-equilibrium processes, Zh. Eksp. Teor. Fiz. 47, 1515 (1965); [JETP 20, 1018 (1965)].
\bibitem{keldysh1}  L. V. Keldysh, “Real-time nonequilibrium green’s functions,” in Progress  in  Nonequilibrium  Green’s  Functions II, pp. 4-17 (World Scientific, 2003).
\bibitem{danielewicz} P. Danielewicz, Ann. Phys. 152, 239–304 (1984).
\bibitem{stef van} G. Stefanucci, R. van Leeuwen, Nonequilibrium Many-Body Theory of Quantum Systems: A Modern Introduction; Cambridge University Press: Cambridge, (2013).
\bibitem{stan} A. Stan, N. E. Dahlen, R. van Leeuwen, J. Chem. Phys. 130, 224101 (2009).
\bibitem{bonitz} K. Balzer, S. Bauch, and M. Bonitz, Phys. Rev. A 81, 022510 (2010).
\bibitem{balzer} K. Balzer, M. Bonitz, Nonequilibrium Green’s Functions Approach to Inhomogeneous Systems; Springer, (2012).
\bibitem{lipavsky} P. Lipavsk\'{y}, V. \v{S}pi\v{c}ka, B. Velick\'{y}, Phys. Rev. B 34, 6933–6942 (1986).
\bibitem{latini} S. Latini, E. Perfetto, A.-M. Uimonen, R. van Leeuwen, G. Stefanucci, Phys. Rev. B 89, 075306 (2014). 
\bibitem{perfetto}  E. Perfetto, A.-M. Uimonen, R. van Leeuwen, and G. Stefanucci, Phys. Rev. A 92, 033419 (2015).
\bibitem{numericalrecipes} W. H. Press, S. A. Teukolsky, W. T. Vetterling, and B. P. Flannery, Numerical Recipes 3rd Edition: The Art of Scientific Computing (Cambridge University Press, New York, 2007), ISBN 0521884071.
\bibitem{Jitomirskaya1999} S. Y. Jitomirskaya, Ann. of Math. 150, 1159 (1999).

\end{thebibliography}
\end{document}